\documentclass[a4paper,11pt]{article}
\usepackage{pos}

\usepackage{latexsym}
\usepackage{hyperref}
\usepackage{graphicx}
\usepackage{epstopdf}
\usepackage{bm}
\usepackage{tikz-feynman}
\title{Nucleon isovector form factors from physical-mass 2+1-flavor dynamical domain-wall QCD}
\ShortTitle{DWF nucleon form factors}

\author*{Shigemi Ohta}

\affiliation{Theory Center, Institute of Particle and Nuclear Studies, KEK\\ 
Oho 1-1, Tsukuba, 3050801, Japan}

\emailAdd{shigemi.ohta@kek.jp}

\abstract{\vspace{-152mm}\parbox{\textwidth}{\flushright\large\rm \hfill KEK-TH-2359}\vspace{145mm}
The current status is reported of joint lattice numerical calculations by LHP and RBC collaborations of isovector nucleon form factors using 2+1-flavor physical-mass domain-wall fermions ensemble at a lattice cutoff \(a^{-1}\) of 1.730(4) GeV and spatial volume of \((La=\mbox{\rm 5.471(13)fm})^3\) generated jointly by RBC and UKQCD collaborations.}

\FullConference{%
 The 38th International Symposium on Lattice Field Theory, LATTICE2021
  26th-30th July, 2021
  Zoom/Gather@Massachusetts Institute of Technology
}

\begin{document}
\maketitle


RIKEN-BNL-Columbia (RBC) and UKQCD collaborations have been jointly generating dynamical 2+1-flavor domain-wall fermions (DWF) numerical lattice-QCD ensembles for over a decade now \cite{RBC:2006jmm,RBC:2007yjf,RBC:2007bso,RBC:2008cmd,RBC-UKQCD:2008mhs,RBC:2010qam,Aoki:2010pe,RBC:2012cbl,RBC:2014ntl,Boyle:2015exm}.
We have been working at essentially physical mass for a while \cite{RBC:2014ntl,Boyle:2015exm}.

\begin{figure}[b]
\begin{center}
\includegraphics[clip,width=.9\textwidth,]{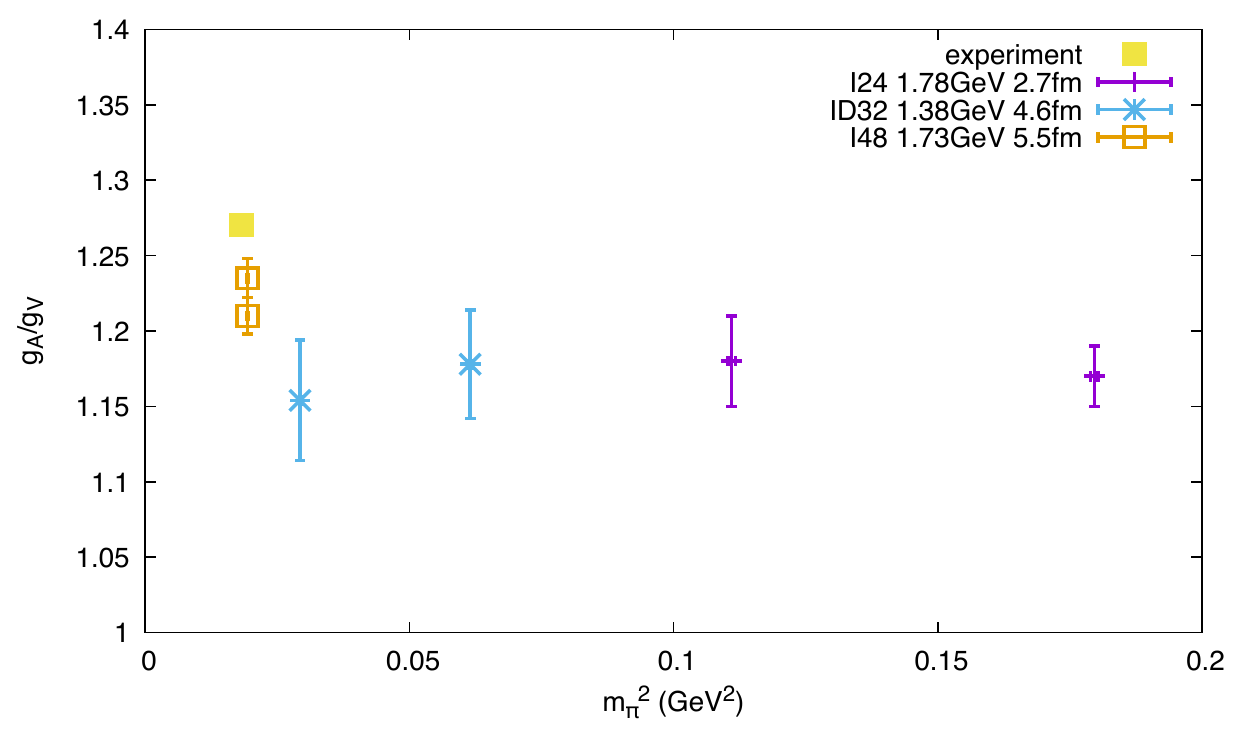}
\caption{
\label{fig:mpi2AV}
The ratio, \(g_A/g_V\), of nucleon isovector axial charge, \(g_A\), to vector charge, \(g_V\),  calculated on successive RBC+UKQCD 2+1-flavor dynamical DWF ensembles with pion mass from 416.4(1.2) and 329.4(1.3) MeV (two righter most points), 249.4(3) and 172.3(3) MeV (two middle points),  and at essentially physical mass of 139.2(2) MeV (two left points just below the experiment), compared with the experimental value (left most).
Earlier and heavier four points show about ten percent deficit from the experiment which agreed with
almost all other lattice numerical calculations at similar mass and cut off.
Though the physical mass result approaches the experiment, we still see significant deficit.
}
\end{center}
\end{figure}

We have used some of these DWF ensembles for studying nucleon \cite{Yamazaki:2008py,Yamazaki:2009zq,Aoki:2010xg,Ohta:2011vv,Lin:2011vx,Lin:2012nv,Ohta:2013qda,Ohta:2014rfa,Syritsyn:2014xwa,Ohta:2015aos,Abramczyk:2016ziv,Ohta:2017gzg,Ohta:2018zfp,Ohta:2019tod,Abramczyk:2019fnf}.
We found deficit \cite{Yamazaki:2008py} in calculated isovector axial charge, \(g_A\), in comparison with its experimental value \cite{10.1093/ptep/ptaa104}.
As we refined our analysis with lighter-mass ensembles (see Fig.\ \ref{fig:mpi2AV})
about ten percent deficit of the calculated results with pion mass from about 420 MeV to 170 MeV had not moved much \cite{Ohta:2011vv,Lin:2011vx,Lin:2012nv,Ohta:2013qda,Ohta:2014rfa,Ohta:2015aos,Abramczyk:2016ziv,Ohta:2017gzg,Abramczyk:2019fnf}.
This was confirmed by almost all other calculations at similar lattice cuts off and quark mass \cite{Dragos:2016rtx,Bhattacharya:2016zcn,Liang:2016fgy,Ishikawa:2018rew,Chang:2018uxx}. 
Since then more calculations at almost physical mass have been conducted, bringing the calculated values to closer to the experiment \cite{Chang:2018uxx,Bhattacharya:2016zcn,Shintani:2018ozy,Hasan:2019noy,Harris:2019bih}, sometimes covering the experimental value within relatively large statistical and systematic errors.

However our unitary DWF calculations with better chiral and flavor symmetries observe some deficit with much smaller statistical errors:
statistiical significance of these results range from three to five standard deviations dependent on renormalization methods \cite{Ohta:2018zfp,Ohta:2019tod}.

As the corresponding vector charge calculation suggests possible contamination from near by excited states \cite{Ohta:2018zfp,Ohta:2019tod}, in contrast to earlier DWF calculations that did not find any evidence for such contamination \cite{Abramczyk:2019fnf}, the form factor calculations presented here can shed some light on this possible contamination from excited states \cite{Bar:2021crj,Bar:2021zds}.


The results presented here were calculated using the ``48I'' \(48^3\times 96\) 2+1-flavor dynamical M\"{o}bius DWF ensemble at physical mass with Iwasaki gauge action of gauge coupling, \(\beta=2.13\), or of lattice cut off of  \(a^{-1} = 1.730(4)\) GeV, jointly generated by the RBC and UKQCD Collaborations \cite{RBC:2014ntl}.
In total 130 configurations, separated by 20 MD trajectories in the range of trajectory number 620 to 980, and by 10 MD trajectories in the range of trajectory number from 990 to 2160, except the missing 1050, 1070, 1150, 1170, 1250, 1270, and 1470, are used.
Each configuration is deflated with 2000 low Dirac eigenvalues \cite{Clark:2017wom}.
The ``AMA'' statistics trick  \cite{Shintani:2014vja}, with \(4^4=256\) AMA sloppy samples unbiased by 4 precision ones from each configuration, is used.
Gauge-invariant Gaussian smearing  \cite{Alexandrou:1992ti,Berruto:2005hg} with similar parameters as in the past RBC nucleon structure calculations is applied to nucleon source and sink, separated by \(8 \le T \le 12\) in time.

\begin{figure}[b]
\begin{center}
\includegraphics[clip,width=.9\textwidth,]{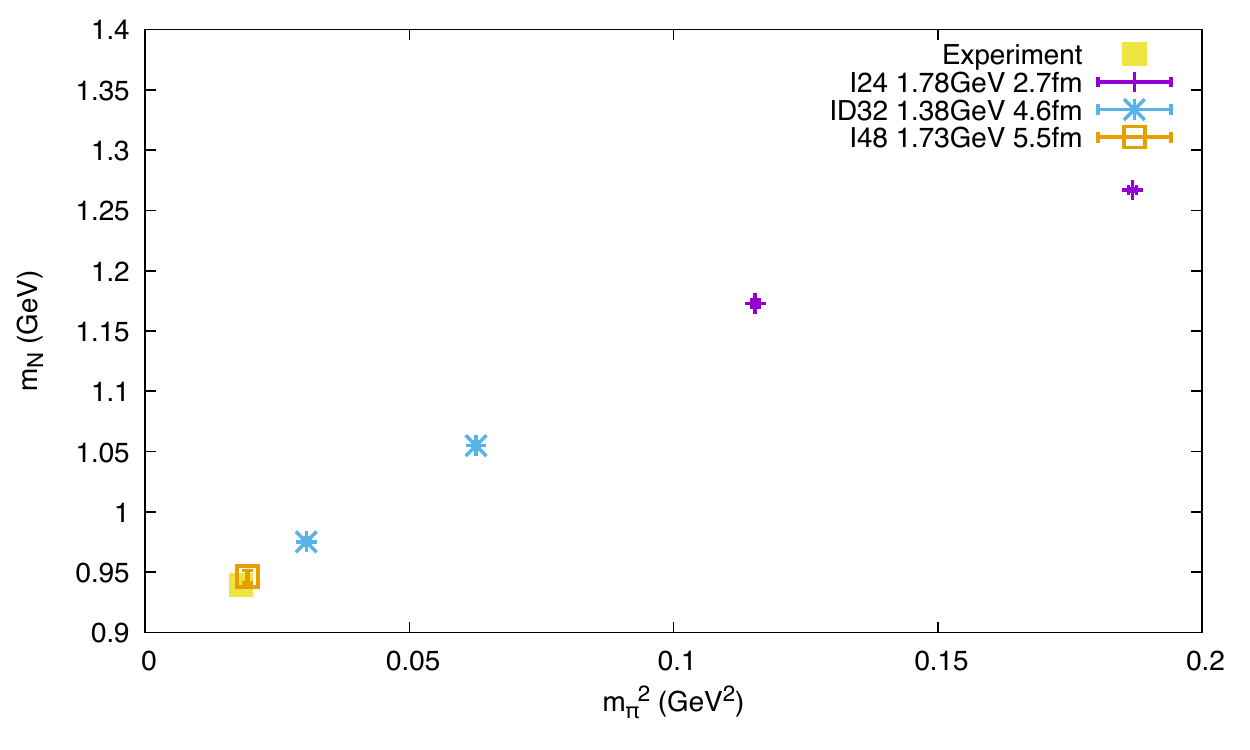}
\caption{
\label{fig:mpi2mN}
Calculated nucleon mass, \(m_N\), plotted against pion mass squared, \(m_\pi^2\), in physical units, from recent RBC+UKQCD 2+1-flavor dynamical numerical lattice QCD ensembles.
The calculated mass appears convex upwards, suggesting possible chiral logarithm.
}
\end{center}
\end{figure}

We obtained a nucleon mass estimate of 947(6) MeV \cite{Ohta:2018zfp,Ohta:2019tod}.
In Fig.\ \ref{fig:mpi2mN} we present the calculated nucleon mass along with four earlier RBC+UKQCD calculations.
Though the calculations were done with different gauge actions and quark mass and are yet to be taken to the respective continuum limits, we see a convex upward quark-mass dependence that trends to the physical nucleon mass.
This likely is a result of much expected chiral logarithm, \(m_\pi^2 \log m_\pi^2 \sim m_{u+d} \log m_{u+d}\).
Note such chiral effects can affect nucleon charges and couplings as well as the mass through for example Goldberger-Treiman relation, \(m_N g_A \propto g_{\pi NN} f_\pi\) \cite{Goldberger:1958tr}.


\begin{figure}[b]
\begin{center}
\includegraphics[width=.9\textwidth,clip]{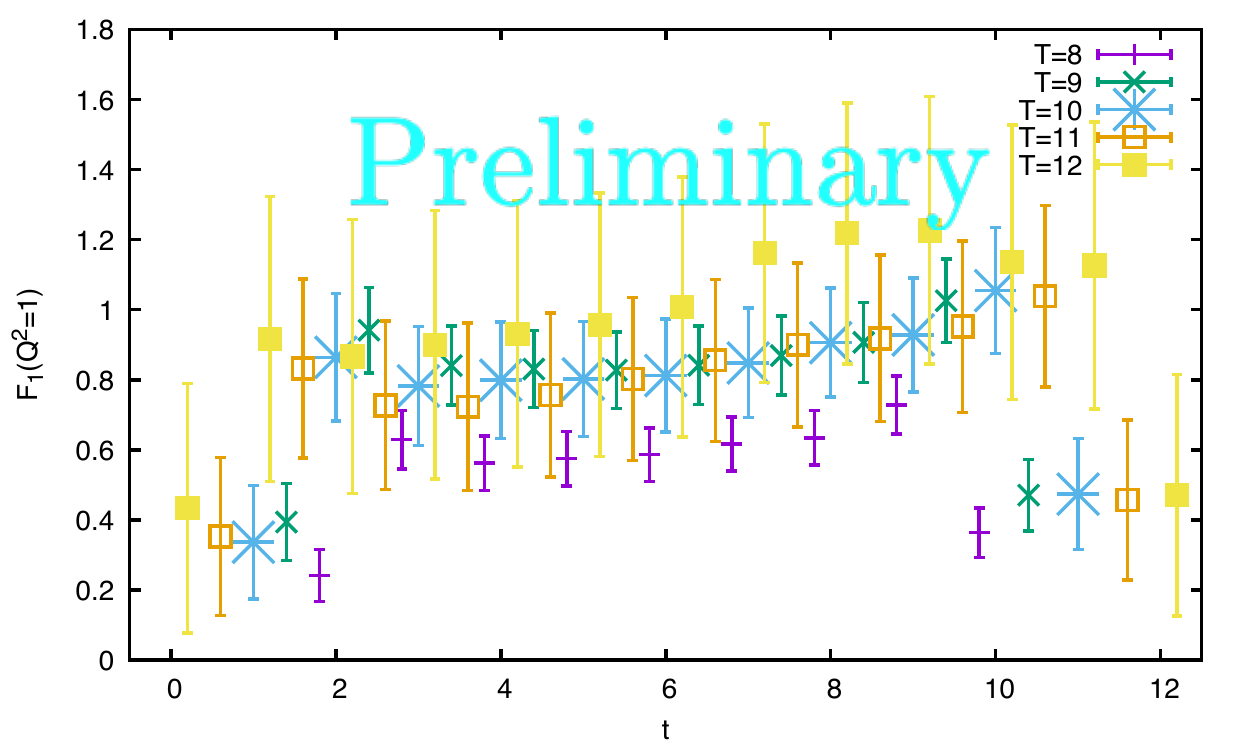}
\caption{
\label{fig:F1}
Nucleon isovector vector form factor \(F_1\) with one lattice unit of momentumn transfer squared, \(Q^2=1\), plotted against source-sink separations, \(T\), of 8, 9, 10, 11, and 12 lattice units.}
\end{center}
\end{figure}
We are presently trying to extract nucleon isovector form factors, which are experimentally measured in elastic scatterings off, or  \(\beta\) decay of, or muon capture by nucleons:
\[\langle p| V^+_\mu(x) | n \rangle = \bar{u}_p \left[\gamma_\mu
F_1(q^2) - i \sigma_{\mu \lambda}q_{\lambda} \frac{F_2(q^2)}{2m_N} \right]
u_n e^{iq\cdot x},
\]
\[
\langle p| A^+_\mu(x) | n \rangle = \bar{u}_p
            \left[\gamma_\mu \gamma_5  F_{A}(q^2)
             +\gamma_5 q_\mu \frac{F_{P}(q^2)} {2m_N}\right]  u_n e^{iq\cdot x}.
\]
Some alternative combinations of these form factors, such as \(\displaystyle
F_V=F_1, F_T=F_2; G_E=F_1-\frac{q^2}{4m_N^2}F_2, G_M=F_1+F_2,
\)
also appear in the literature.
They are related to various important nucleon observables such as: mean-squared charge radii, \(\langle r_E^2\rangle\), through the expansion of the vector form factor, \(\displaystyle F_1 = F_1(0) - \frac{1}{6} \langle r_E^2\rangle Q^2 + ...\), in terms of momentum transfer squared, \(Q^2 = |q^2|\),  or anomalous magnetic moment,  \(F_2(0)\), or isovector axial charge, \(g_A=F_A(0)=1.2754(13) g_V\)  \cite{10.1093/ptep/ptaa104}, of nucleon that determines neutron life and nuclear \(\beta\) strengths that in turn determines nuclear abundance.

We use the standard three-point to two-point correlator ratios to extract the form factors, the details of which can be found in our earlier publications such as Ref.\ \cite{Yamazaki:2009zq}:
after appropriate projections and normalization, we obtain the relevant plateaux between the nucleon source and sink (Fig.\ \ref{fig:F1}).
As in the earlier studies for isovector charges and couplings \cite{Ohta:2018zfp,Ohta:2019tod} we use the source-sink separation, \(T\), from 8 to 12.

The preliminary results shown in Fig.\ \ref{fig:F1} are rather noisy even for the smallest finite momentum transfer squared, \(Q^2=1\) lattice unit.
Though statistically not significant through the large noise, the results from the shortest source-sink separation may be systematically lower than the others.
This could mean steeper slope momentum transfer squared, or larger mean-squared charge radius, at shorter source-sink separation.
Contamination from excited-state that is larger in mean-squared charge radius than the ground state can cause such an effect.

Here again in isovector form factors, just like it would have in the cases of isovector charges  \cite{Ohta:2018zfp,Ohta:2019tod}, it would help to obtain more statistics to better understand their behaviors:
Indeed doubling the present statistics could clarify the question of excited-state contamination.
Adding a couple of shorter source-sink separations such as \(T=7\) and 6 would be helpful too.


Doubling our statistics could resolve the question about possible excited-state contamination, not only in the form factors presented here but also in the charges reported earlier.
It would also help to add a couple of more shorter source-sink separations such as \(T=7\) and 6 to understand their behaviors better.
Given the present three to five standard-deviation statistical significance  \cite{Ohta:2018zfp,Ohta:2019tod}, the deficit in isovector axial charge, \(g_A\), is likely to persist.
If it does it would be interesting to study the effects of isospin violation, both from the mass and electric charge differences.


The author thanks the members of LHP, RBC, and UKQCD Collaborations, and in particular Sergey Syritsyn.
The ``48I'' ensemble was generated using the IBM Blue Gene/Q (BG/Q) ``Mira'' machines at the Argonne Leadership Class Facility (ALCF) provided under the Incite Program of the US DOE, on the ``DiRAC'' BG/Q system funded by the UK STFC in the Advanced Computing Facility at the University of Edinburgh, and on the BG/Q  machines at the Brookhaven National Laboratory.
The nucleon calculations were done using ALCF Mira.
The author was partially supported by Japan Society for the Promotion of Sciences, Kakenhi grant 15K05064.
Part of the work was conducted while the author was affiliated with RIKEN-BNL Research Center through March 31, 2021.

\bibliographystyle{apsrev4-2}
\bibliography{nucleon}

\end{document}